# Incorporating LLMs for Large-Scale Urban Complex Mobility Simulation

Yu-Lun Song[1], Chung-En Tsern[3], Che-Cheng Wu[2], Yu-Ming Chang[2],
Syuan-Bo Huang[2], Wei-Chu Chen[2], Michael Chia-Liang Lin[1], Yu-Ta Lin[2]

[1]Media Lab @ Massachusetts Institute of Technology
[2]City Science Lab @ National Taipei University of Technology
[3]University College London

**Summary**

This study presents an innovative approach to urban mobility simulation by integrating a Large Language Model (LLM) with Agent-Based Modeling (ABM). Unlike traditional rule-based ABM, the proposed framework leverages LLM to enhance agent diversity and realism by generating synthetic population profiles, allocating routine and occasional locations, and simulating personalized routes. Using real-world data, the simulation models individual behaviors and large-scale mobility patterns in Taipei City. Key insights, such as route heat maps and mode-specific indicators, provide urban planners with actionable information for policy-making. Future work focuses on establishing robust validation frameworks to ensure accuracy and reliability in urban planning applications.

## 1. Introduction

Mobility reflects the unique geographic, economic, and cultural contexts of cities while being shaped by and confined to the urban infrastructure that supports it. This intricate interplay impacts accessibility, convenience, and quality of life, offering critical insights into how cities function and adapt (Sheller and Urry 2006; Kang et al. 2012). Understanding these dynamics is crucial for urban planning, emphasizing the need for transportation systems that respond to current urban contexts while shaping future mobility patterns. Mobility simulation is a key tool in this process, enabling scenario evaluation, resource optimization, and the design of sustainable urban environments.

Agent-based modeling (ABM) is a prominent approach for urban simulation, as it captures the behaviors of individual agents to study the impact of heterogeneous human behaviors on urban systems. However, traditional ABM often relies on rule-based behaviors, which can oversimplify the complexity of individual actions within urban environments. This simplification may lead to simulation outcomes that are less representative of real-world dynamics (Heppenstall, Malleson, and Crooks 2016). Recent advancements propose integrating ABM with Large Language Models (LLMs) to overcome these limitations. Platforms like OpenCity and the Smart Agent-Based Modeling (SABM) framework demonstrate LLMs' potential to enhance scalability, realism, and complexity in urban simulation (Yan et al. 2024; Wu et al. 2023).

[1]allen017@media.mit.edu
[2]chung-en.tsern.23@ucl.ac.uk
[3]reeve0319@gmail.com
[4]roger@mail.ntut.edu.tw
[5]syuanbo@mail.ntut.edu.tw
[6]csl_drew@mail.ntut.edu.tw
[7]mcllin@mit.edu
[8]roylin@mail.ntut.edu.tw

This research aims to explore the potential of the integration of LLMs with ABM to diversify the simulation and make the outcome be more interpretable.

## 2. Methods

**Figure 1** is the demonstration of the research workflow.

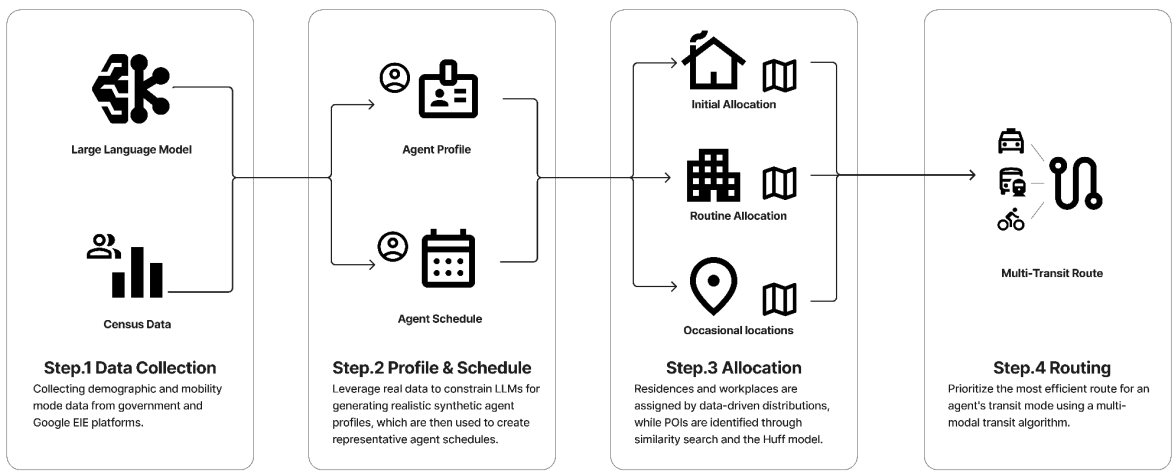

**Figure 1** The workflow of the simulation.

### 2.1. Profile generation

In this study, a large language model (LLM) is utilized as a tool for establishing relationships among diverse statistical data points. De-identified open statistical data—including variables such as age, education level, occupation, salary distribution, and mobility preferences—serve as inputs for the LLM, which models the underlying correlations among these variables to produce a vertically integrated distribution framework. As illustrated in **Figure 2**, the LLM processes statistical data inputs, such as age and education level, to generate proportional distributions for each age group. Iterative Proportional Fitting algorithm ensures that the aggregated educational distribution aligns with real-world population-level statistics. With the LLM's inherent recognition of society and human behaviors, the generated synthetic profiles exhibit consistent and logically coherent attributes. Consequently, the synthetic profiles relatively closely mirror real-world population characteristics while circumventing individual privacy by operating on de-identified statistical data.

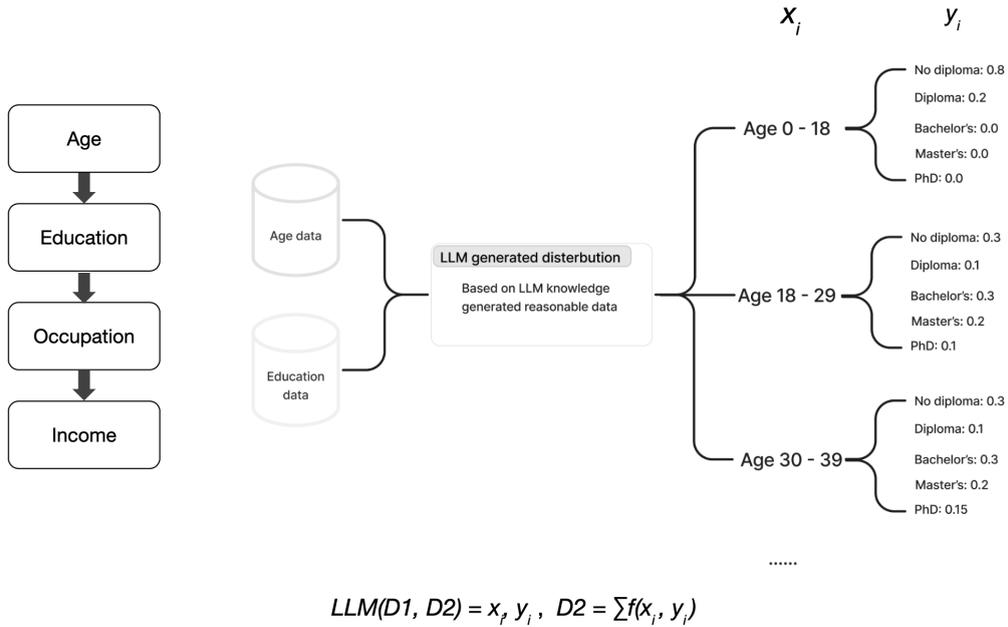

$$LLM(D1, D2) = x_i, y_i, \quad D2 = \sum f(x_i, y_i)$$

**Figure 2** The illustration of a LLM establishes relationships between statistical data while the aggregated distribution aligns with the original data distribution.

## 2.2. Allocation

The city is divided into uniformly sized grids of 250 meters by 250 meters to model the spatial distribution of individuals within the urban environment. Each grid's population capacity is determined based on a combination of census data and average income level data. This information guides the allocation of agents' initial and routine locations. Agents' initial locations are determined by matching their income to the average income level of the corresponding grid, as illustrated in **Figure 3**. For routine locations, the LLM is employed to generate descriptions of an agent's occupation and relevant industrial categories, to increase more information. By performing text similarity matching, each occupation is mapped to its corresponding industry, shown in **Figure 4**. Routine locations are then assigned by randomly selecting a point within the matched industry category from the available Points of Interest (POI) dataset.

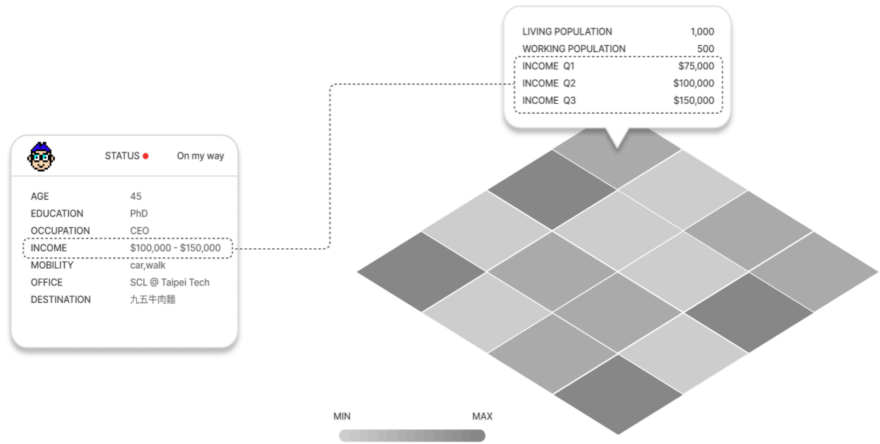

**Figure 3** The allocation of initial location based on average income level data.

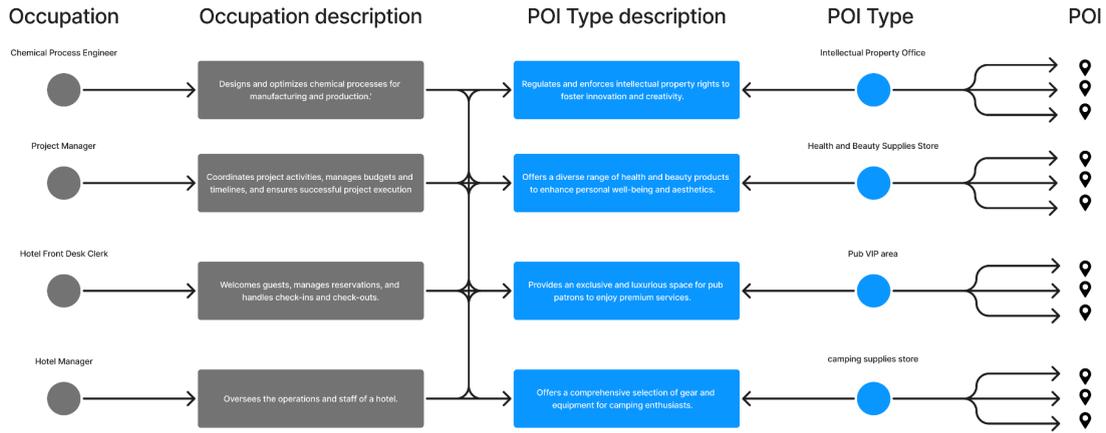

**Figure 4** The workflow of the allocation of the routine location.

### 2.3. Occasional location matching

To reflect the dynamics of a city as much as possible, occasional locations are modeled by referring to the schedule of agents generated by LLM. Each activity in the schedule generated by LLM referring to the agent's profile is matched to a corresponding Point of Interest (POI) category through semantic similarity-based mapping. By referring to the Huff model (Garcia-Gabilondo, Shibuya, and Sekimoto 2024), the final location is selected using the modified version (**Equation 1**) that considers both distance and attractiveness, where attractiveness is a composite score of popularity and credibility. A weight is assigned to each candidate location based on these factors, and the occasional location is chosen probabilistically, maintaining flexibility while reflecting realistic spatial patterns.

$$weight\ =\ \frac{attractiveness}{distance^{decay}}\ =\ \frac{popularity \times credibility}{distance^{decay}} \qquad (1)$$

### 2.4. Routing

To simulate realistic mobility behavior, the Multi-Criteria Range Raptor (McRAPTOR) algorithm (Delling, Pajor, and Werneck 2015), which accommodates diverse individual preferences, is utilized to generate personalized routing solutions. McRAPTOR iteratively evaluates all possible routes and transfer options across multiple rounds to identify optimal paths that are not outperformed in any single criterion. By incorporating multiple optimization objectives and balancing trade-offs, the algorithm enables the simulation to closely mirror real-world decision-making and transportation behaviors. Additionally, it serves as a crucial tool for calculating key indicators, allowing for a comprehensive assessment of the mobility impact on the simulated environment.

### 3. Results & Discussion

**Figure 5** demonstrates a one-day simulation, approximately 100,000 agents are deployed across Taipei City, Taiwan. Each point on the map represents an agent, with the brightness of the color indicating the density of agents in that area. This comprehensive visualization enables observers or planners to understand agent behavior patterns and environmental impacts during the simulation period.

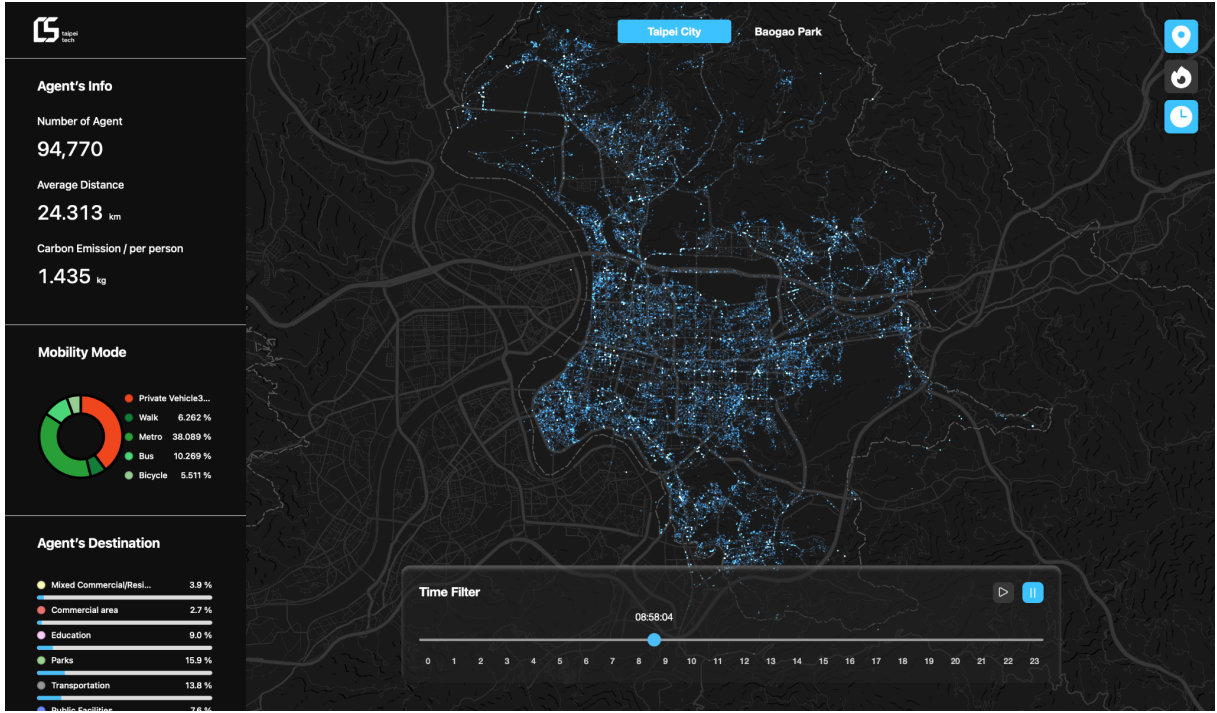

**Figure 5** The platform of the simulation outcome.

By integrating the capabilities of ABM and LLM, the platform allows observers to delve into individual agent profiles and daily schedules, as demonstrated in **Figure 6**. Each agent's trajectory, activity locations, and transportation modes can be tracked and analyzed. This micro-scale observation aids in understanding individual mobility behaviors and the types of people gathered in various locations at different times.

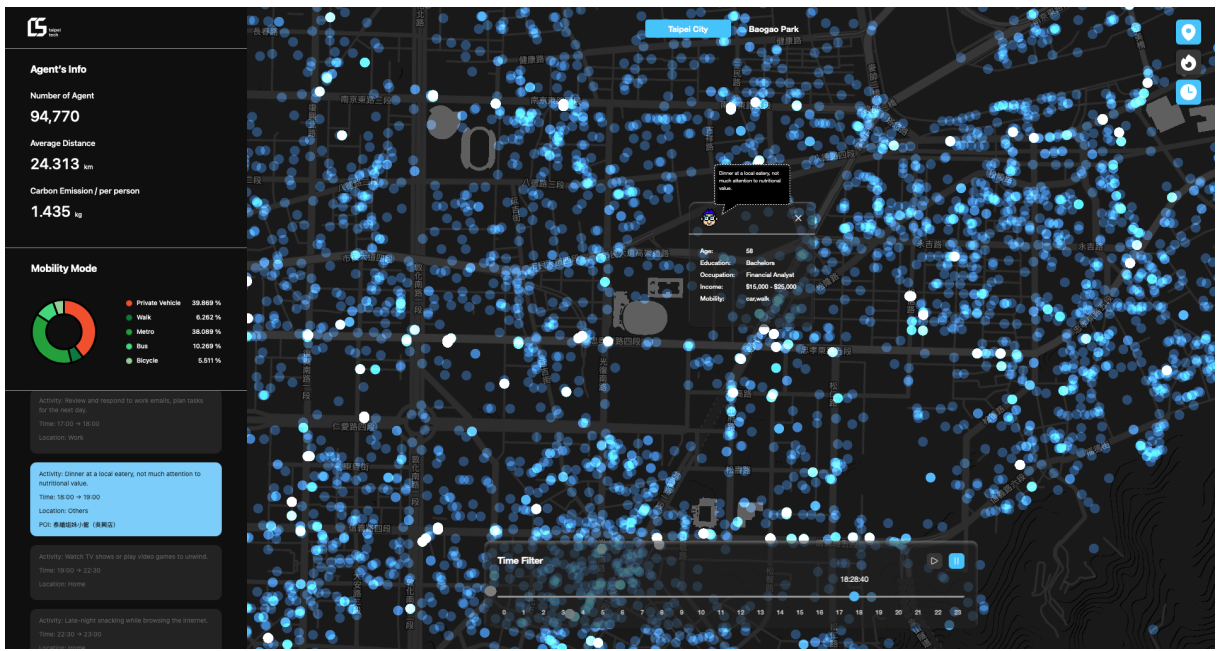

**Figure 6** Persona of an agent and its one-day schedule in the simulation.

In addition to individual-level analysis, it also offers macro-scale insights through route heat maps. **Figure 7** displays the route heat map for private vehicles during the morning period, while **Figure 8** shows the corresponding heat map for pedestrians. By examining these visualizations, urban planners can assess which roads experience higher traffic and determine the dominant transportation modes during specific time intervals. This information is crucial for identifying traffic hotspots and evaluating the effectiveness of existing policies.

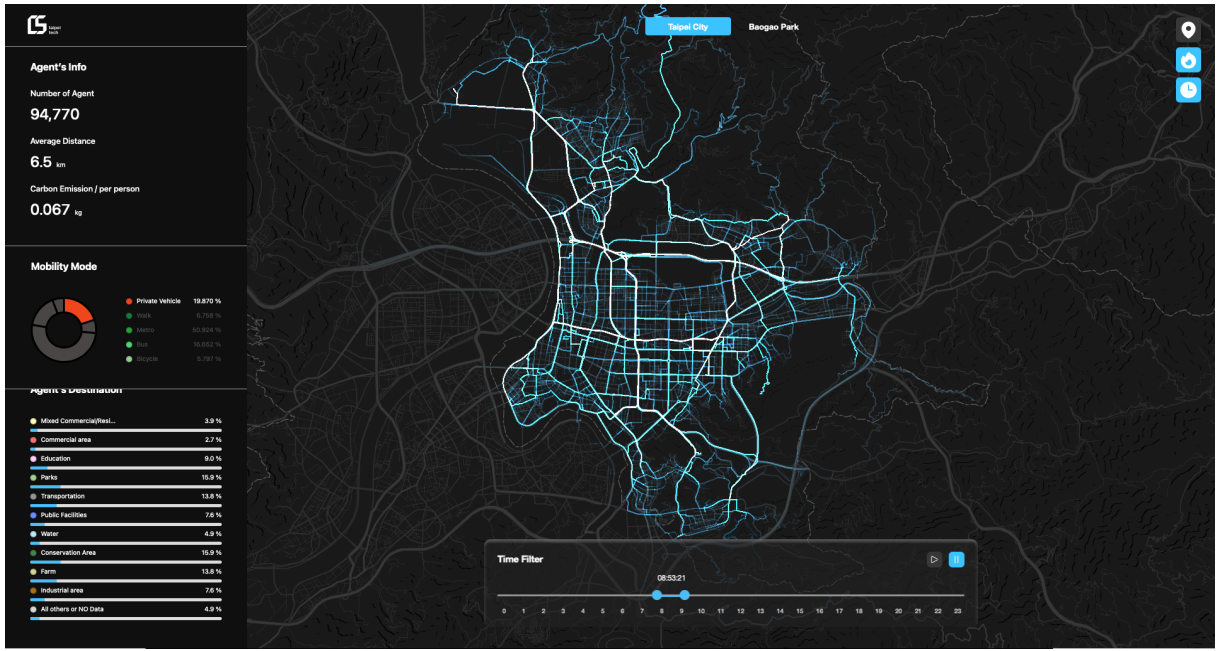

**Figure 7** Route heat map of private vehicles during the morning in Taipei City.

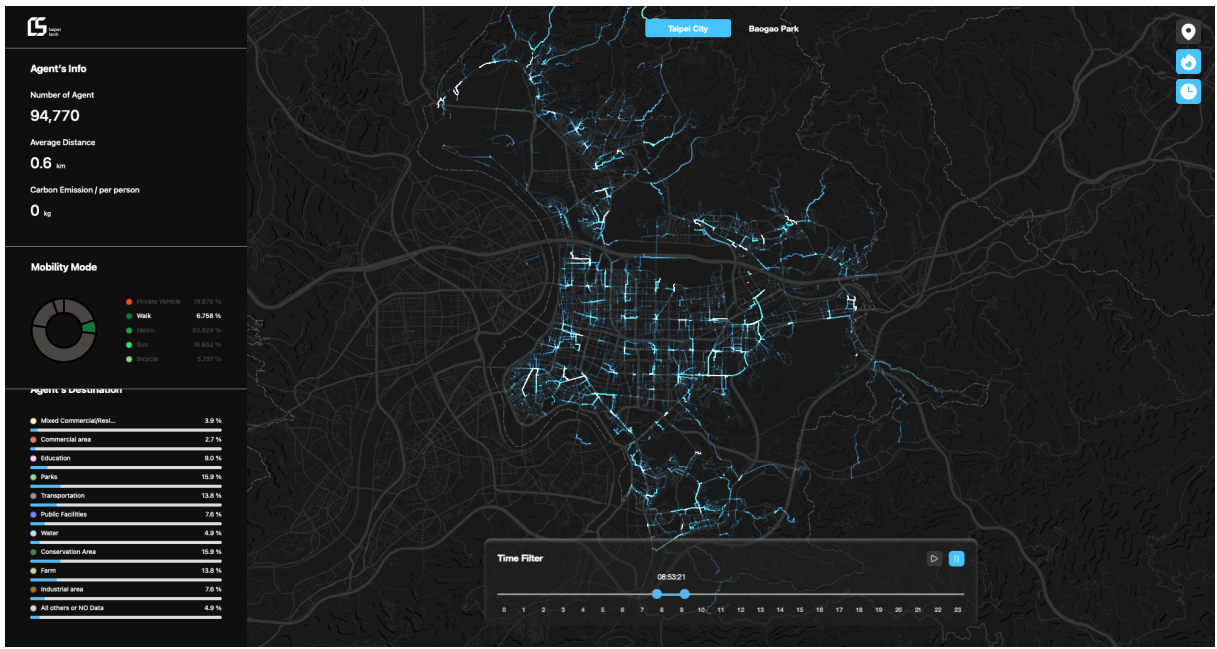

**Figure 8** Route heat map of pedestrians during the morning in Taipei City.

The left-hand panel also presents key indicators, such as the proportion of various mobility modes, average travel distance, and carbon emissions. These metrics provide valuable quantitative insights that can guide policy-making. With access to both micro-scale agent profiles and macro-scale route patterns, urban planners and researchers can leverage this simulation to derive actionable insights for future urban development.

## 4. Conclusion

This study explores the potential of leveraging the capability of LLM with ABM to conduct more diverse and interpretable urban simulations. The simulation results indicate significant promise in providing urban planners with valuable insights for future city development. Nevertheless, the accuracy of these results requires rigorous verification to ensure reliability. Future research should focus on establishing robust validation frameworks to assess the accuracy of each model component. Achieving this objective will necessitate comprehensive data collection from local urban environments to ensure sufficient coverage and representativeness. Ultimately, this approach will enhance the credibility and applicability of simulation-driven urban planning.

**Biographies**

Yu-Lun Song is a graduate student at the MIT Media Lab and a research assistant in the City Science Lab, specializing in AI and urban mobility simulation.

Chung-En Tsern is a graduate student at UCL CASA and a research assistant at the City Science Lab @ Taipei Tech, specializing in spatial data, policy research, and urban simulation.

Che-Cheng Wu, Research Assistant at the City Science Lab @ Taipei Tech, focused on utilizing large language models (LLMs) for urban data analysis and simulation.

Yu-Ming Chang, Researcher at City Science Lab @ Taipei Tech, specialized in data analysis and transforming urban data research into applications.

Syuan-Bo Huang, Researcher at the City Science Lab @ Taipei Tech, dedicated to leveraging the application of AI and urban data in solving intricate urban issues.

Wei-Chu Chen, Researcher, focuses on spatial analysis and routing tools to address complex urban challenges.

Dr. Michael Lin is Research Scientist at MIT and Director of Taipei City Science, specializing in complex urban systems with over 15 years of experience in leading collaboration with Fortune 500 & the public sectors around the world to craft future urban mobility from concepts to reality.

Yu-Ta Lin, Head of Urban Informatics, City Science Lab @ Taipei Tech, specialized in urban data analysis for urban design and planning.